\documentclass[12pt]{article}

\date{}

\begin{document}
\thispagestyle{empty}
\title{\bf A possible explanation why the $\Theta^+$ is seen \\
in some experiments and not in others}
\author{Ya. Azimov$^{1}$, K. Goeke$^2$, I. Strakovsky$^3$\\ \\
$^1$ Petersburg Nuclear Physics Institute,\\ Gatchina, 188300 Russia \\
$^2$ Institut f\"ur Theor. Physik -II, Ruhr-Universit\"at,\\
D-44780 Bochum, Germany\\
$^3$ Center for Nuclear Studies, Physics Department, \\ The George
Washington University,\\ Washington, DC 23606 USA} \maketitle

\begin{abstract}
To understand the whole set of positive and null data on the
$\Theta^+(1530)$-production, we suggest the hypothesis that
multiquark hadrons are mainly generated from many-quark states,
which emerge either as short-term hadron fluctuations, or as
hadron remnants in hard processes. This approach allows us to
describe both non-observation of the $\Theta^+$ in current null
experiments and peculiar features of its production in positive
experiments. Further, we are able to propose new experiments that
might be decisive for the problem of the $\Theta^+$ existence.
Distributions  of the $\Theta^+$ in such experiments can give
important information both on higher Fock components of
conventional hadrons and about structure and hadronization
properties of hadron remnants produced in hard processes. We also
explain that description of multiquark hadrons may require a
modified form of the constituent quark model, with quark masses
and couplings being intermediate between their values for the
familiar constituent quarks and the current ones.


\end{abstract}

\section{Introduction}

Invented more than 40 years ago~\cite{GN,GM,Zw}, quarks were initially
introduced rather formally, to account for a very limited variety
of hadronic flavor multiplets known at that time. Their simplest
application was to present every baryon as a three-quark system
and every meson as a quark-antiquark system. Then, in respect to the
flavor symmetry group $SU(3)_F$, the mesons could exist only as
singlets and/or octets, while the baryons could reveal both the same
kinds of multiplets and, in addition, decuplets.

Such a simple picture was in good correspondence with experiment.
It could be quite successful, if the quarks were nothing more than
mathematical objects allowing to visualize symmetry classification
of hadrons (such treatment would not contradict to Gell-Mann's
paper~\cite{GM}). However, if quarks (or ``aces'', in Zweig's
terminology~\cite{Zw}) are physical objects, the picture of hadrons
with a fixed number of constituents could be self-consistent only
in a non-relativistic theory. In a relativistic case, production and
annihilation of quark-antiquark pairs prevent the total number of quarks
and antiquarks from being fixed.  That is why the first hints, that the
quarks might be something more than just mathematical objects, revived
the question whether one would be able to observe other (exotic) kinds
of flavor multiplets, with some more complicated quark content.

One of the simplest clear examples of non-3-quark baryons would be
a baryon with positive strangeness, say, a resonance in the $KN$
system with $S=+1$ (by definition, the strange quark $s$ has
$S=-1$, and a 3-quark system cannot have $S=+1$). Experimental
searches for such exotic states in the $KN$-spectra started rather
early~\cite{cool,abr,tys}. However, all suggested evidences stayed
unconvincing, and later the Particle Data Group (PDG) stopped
discussions on experimental spectroscopy of exotics~\cite{PDG86}.

The first theoretical attempt to describe internal dynamical
structure for specific multiquark hadrons was made in the
framework of the ``MIT bag"~\cite{jaf1,jaf2,jaf3}. The
calculations supported existence of exotic states, but prescribed
them to be very broad, with widths of some hundreds MeV. This
result seems quite understandable, since an exotic multiquark
system in the MIT bag looks to be readily prepared for separation
into subsystems with conventional quark contents: a tetraquark
meson $qq\overline{q}\overline{q}$ may be considered as a system
of two quark-antiquark pairs, while the pentaquark baryon
$qqqq\overline{q}$ may be considered as (3 quarks + 1
quark-antiquark pair). According to the MIT-bag approach, it is
just the enormously huge width that could explain why exotic
resonances have not been seen in experiment: they exist, but are
too broad to reveal clear-cut bumps. Totally unnoticed for long
time has stayed an alternative possibility: the fact that hadrons
have an internal structure may suggest some hadrons to have a very
complicated structure which would suppress their couplings (and,
therefore, production cross sections and decay widths) to
conventional hadrons~\cite{az70}. It could be similar to
``structural'' suppression of some radiative transitions, which is
well-known in atomic physics. Such possibility has been recently
demonstrated by detailed calculations for the system of
$4q\overline{q}$~\cite{cc}.

It is interesting, that the standard partial-wave analysis of the
elastic $KN$-scattering~\cite{hys} (the latest and most complete
one published up to now) has later presented four exotic baryon
states (resonances), two isoscalar and two isovector, all they
having, indeed, large widths 200--500~MeV. However, even before
the publication~\cite{hys}, any correspondence between the MIT bag
and experiment had become to look generally dubious.

New impetus for studies of exotics has emerged~\cite{dpp} from the
Chiral Quark Soliton Approach ($\chi$QSA; for recent reviews see
Refs.\cite{wk,ekp}). It allowed to give rather detailed
theoretical prediction of the exotic antidecuplet of baryons
(including the $\Theta^+$ with $S=+1$) and strongly stimulated new
experimental searches for exotics. These efforts gave, at last,
some positive evidences, as summarized in the Review of Particle
Properties, issue of 2004~\cite{PDG04}.

Nevertheless, the current experimental status of the exotic baryon
$\Theta^+(1530)$ is rather uncertain (see the more recent
experimental review~\cite{bur} and the latest edition of the
Review of Particle Properties~\cite{PDG06}). Some collaborations
give positive evidences for its observation, some others do not
see it, thus casting doubts on existence of the $\Theta^+$ and on
correctness of its positive evidences. Especially impressive are
the recent high-statistics null results of the CLAS Collaboration
for $\Theta^+$-photoproduction in several
final-states~\cite{clas1, clas2,clas3,clas4}. And yet, new
(``after-CLAS'') dedicated analyses have again provided both
confirming~\cite{leps,svd,svd-m,bar,tr} and null~\cite{foc, l3,
ank,tof,nom} publications. Moreover, a new suggested
way~\cite{adp} for data analysis may reveal presence of the
$\Theta^+$ even in the published CLAS results, through its
interference with a known resonance.

At present, an important fact is that there are no data sets, from
independent groups, with exactly overlapping observational
conditions (initial and/or final states, kinematical regions).
Therefore, when comparing today different current data on the
$\Theta^+$, one needs some theoretical models/assumptions for the
unknown production mechanisms.  Thus, one cannot yet reject
existence of the $\Theta^+(1530)$ as a particular exotic
pentaquark baryon. We emphasize also that, up to now, there has
not been suggested, in the framework of Quantum Chromodynamics
(QCD) or in other terms, any theoretical reason to forbid
multiquark hadrons.

In such uncertain situation, Karliner and Lipkin~\cite{kl,lip}
raised the question shown in the title here, why the $\Theta^+$ is
seen in some experiments and not in others. As an answer, they
suggested an important role of the baryon resonance with hidden
strangeness $N(2400)$ (see also Ref.\cite{as}). However, this
resonance production seems to be insufficient today to explain
many of positive evidences. An opposite viewpoint is that all
positive results might arise as statistical fluctuations and do
not reveal a true physical effect (see, \textit{e.g.},
Ref.\cite{hic}).

In the present paper, we assume that the $\Theta^+$, as a
representative of exotic hadrons, does exist and have properties
corresponding to the published positive evidences: rather low mass
and unexpectedly narrow width. Then we are going to reconsider
possible dynamical picture for multiquark hadrons. Though our
approach is still qualitative, not quantitative yet, it seems to
explain some regularities in (non)observation of the $\Theta^+$,
It allows, further, to suggest new experiments for more detailed
studies of the $\Theta^+$-production properties. At the moment, we
do not discuss other possible multiquark hadrons, which existence
has today much weaker experimental support than for the
$\Theta^+$.

\section{Exotics and hadron structure}

The latest experiments on exotic hadrons (baryons, really) have
been oriented on predictions coming mainly from $\chi$QSA), first
of all, from Ref.\cite{dpp}. The corresponding models, as well as
other chiral models of strong interactions, may be formally
derived in the framework of QCD after functional integration over
quark and gluon degrees of freedom. Physically, such a procedure
corresponds to averaging over parton (quark and gluon)
distributions inside hadrons, and may provide rather good
description of many properties of hadrons and/or their
interactions. It seems to be true, indeed, for the basic octet and
decuplet baryons (see, \textit{e.g.}, Refs.\cite{chr1,chr2,dia,
dp}). But, as every averaging, it may appear to be insufficient
(or even inadequate) for some problems. Interaction properties of
multiquark hadrons possibly demonstrate just examples of such a
kind.

The $\chi$QSA has provided evidence that some coupling constants
of exotic hadrons should be suppressed~\cite{dpp,ekp,pra}, but has
not explained a physical reason for the suppression. Nevertheless,
traditional calculational methods of $\chi$QSA, related to
decomposition in $1/N_c$, allow to find upper bounds for those
coupling constants. As a result, such methods predict
$\Gamma_{\Theta^+} \leq15~{\rm MeV}$~\cite{dpp}, but do not suggest
any reliable way to find the corresponding numerical value of
$\Gamma_{\Theta^+}$.  Meanwhile, even the upper bound seems to be
unexpectedly low in comparison with familiar widths of baryonic
resonances ($\sim50 - 100$~MeV). But if the $\Theta^+$ does exist,
its experimental width~\cite{bar} \begin{equation} \label{expG}
\Gamma_{\Theta^+}=(0.36\pm0.11)~{\rm MeV} \end{equation} is
strongly below the theoretical $\chi$QSA upper bound~\footnote{The
``model-independent approach'' to the chiral quark soliton, using
phenomenology of the proton spin content and hyperon semileptonic
decays, has recently shown~\cite{mia} that the $\chi$QSA may be
consistent with $\Gamma_{\Theta^+} < 1$~MeV and even with the
value~(\ref{expG}).  However, this approach cannot yet fix a particular
theoretical value for the width.}. More conservative phenomenological
estimations also give only upper bounds for $\Gamma_{\Theta^+}$, which
are $1-5~{\rm MeV}$~\cite{arn,ct,gib,sib}. These values are also
essentially lower than the $\chi$QSA limit, though still much higher
than the strikingly low value~(\ref{expG}).

In such cases one should, evidently, put off the total averaging
over quark and gluon degrees of freedom (which leads to the
$\chi$QSA) and return (at least, for some time) to the original
non-averaged picture. According to the QCD, any hadron should be
described as a Fock column, which components may contain any
number of (additional) quark-antiquark pairs and/or gluons.
Similar is the structure of any bound state in the framework of a
relativistic quantum field theory. The hydrogen atom,
\textit{e.g.}, as described by the Quantum Electrodynamics (QED),
does contain, together with the conventional and familiar
proton-electron configuration, some admixture of virtual photons
and electron-positron pairs. It is the small value of
$\alpha\approx1/137$, which makes the admixture so minute, that
only tiny specific effects are sensitive to its presence. In
difference, the QCD coupling ``constant'' $\alpha_s$ may achieve
the strong interaction regime, and the role of the additional
pairs may become essential in some situations.

Sure, higher Fock components reveal themselves in short-term
fluctuations and, \textit{e.g.}, in hard processes. But one cannot
separate a particular higher component to study it in detail.
Multiquark hadrons, if they exist, could be very helpful in this
relation. Study of pentaquark baryons, having the 5-quark
configuration as a basic component, might help to understand also
properties of the 5-quark component in conventional baryons, where
this component has a less apparent role.

A possible similarity between higher Fock components of
conventional hadrons, on one side, and basic components of
multiquark hadrons, on the other side, would help to
understand/predict the structure of multiquark hadrons. Vice
versa, it may be checked by studying the multiquark hadron
properties. For example, if the basic configurations for
multiquark hadrons are contained in the same confinement area as
for conventional 3-quark hadrons, then the mean distance between
quarks is smaller inside multiquarks. Due to asymptotic freedom of
QCD, this implies that the effective coupling constant (a running
quantity!) is smaller for multiquark hadrons than for conventional
ones. Quark masses (also running quantities!) should also be
smaller inside multiquarks than in the familiar constituent quark
model, based on conventional hadrons. This latter point seems to
agree with data. Indeed, with $m_{u,d}\sim 300$~MeV and
\mbox{$m_s\sim 450$~MeV}, the standard constituent quark model,
applied to the $\Theta^+=(udud\overline s)$, would predict the
mass $M_{\Theta^+} \sim1650$~MeV, while the experimental $\Theta^+$
(if it exists, of course) is about $100$~MeV lighter. Thus, the
constituent quark picture, to be applicable for multiquark
hadrons, might have different values of parameters (\textit{e.g.},
masses) than the familiar constituent quark picture for
conventional hadrons \footnote{Relation between exotic hadrons 
and the constituent quark model has been recently discussed 
from another viewpoint by Lipkin~\cite{lip1}.}.

Note that the QCD Hamiltonian contains only the current quarks; 
thus, the constituent quarks may appear in QCD only as 
effective objects, like, \textit{e.g.}, quasi-particles in the 
solid state physics (even the electron in matter, generally, 
has an effective mass different from its free value). A hint 
that (effective) quark masses may depend on the kind of the 
hadron can be seen even in the early papers of Zeldovich and 
Sakharov~\cite{z&s}. They described masses of hadrons by an 
effective Hamiltonian with several parameters. These parameters 
were fitted separately for the mesons (pseudoscalar and vector) 
and the baryons (octet $1/2^+$ and decuplet $3/2^+$). One of the 
parameters, in framework of the constituent quark model, may be 
treated as the sum of masses of hadron's constituents.  Then, 
results of Zeldovich and Sakharov~\cite{z&s} imply that the 
light quark masses $m_u=m_d$ are 1083/3=361~MeV in baryons, but 
only 598/2=299~MeV in mesons (interestingly, they receive 
nevertheless the same mass difference $m_s-m_u\approx 170$~MeV, 
both in mesons and in baryons).

The change of quark properties for multiquark \textit{vs.} conventional
hadrons looks to be natural. In Fock configurations with infinitely
many quark-antiquark pairs, quarks should be seen as current ones,
with low masses (several MeV for $u$ and $d$) and with $\alpha_s$ in
the asymptotically free regime. Transition to such configurations from
3-quark ones may be expected to reveal some kind of continuity. Then,
quarks in intermediate configurations, and in multiquark hadrons as
well, should have some intermediate properties.

The Fock column picture of hadrons has been recently applied to
the problem of small width of the $\Theta^+$~\cite{dpF}. Starting
from the soliton of $\chi$QSA, the baryons were described at the
light-cone as a set of quarks (3 quarks appended by additional
$q\overline{q}$ pairs) in the self-consistent mean chiral field. Thus,
the gluon degrees of freedom were totally integrated out, while the
quark ones were integrated out only partly. The calculations gave
indeed a small width $\Gamma_{\Theta^+}\approx4$~MeV, essentially
below the $\chi$QSA upper bound, but still higher than the experimental
value (\ref{expG}). The final numerical result is sensitive to
approximations and assumptions. Various corrections may further
diminish the width. Indeed, application of relativistic corrections
gave the width about 2~MeV~\cite{cl,dd}, while further account for
the mass difference between the $\Theta^+$ and nucleon have lowered
the theoretical value down to $\Gamma_{\Theta^+}\approx
0.43$~MeV~\cite{cl1}, just comparable to the experimental value
(\ref{expG}).

Let us discuss the physical reasons for such a small value.
According to the calculations~\cite{dpF,cl,cl1}, emission of the
chiral pseudoscalar field (of kaons for the $\Theta^+$-decay) does
not change the total number of quarks and antiquarks in the
infinite momentum frame. Then, say, the $\Delta(1232)$, being
mainly in a 3-quark configuration, can emit one pion to transform
into the basic 3-quark component of the nucleon. The corresponding
probability is rather high. The same is true for the
$NN\pi$-vertex. In difference, the minimal configuration of the
$\Theta^+$ is 5-quark. After the kaon emission, it transforms into
a non-strange baryon system, but again in the 5-quark
configuration. Therefore, the decay $\Theta^+\to KN$ goes through
multiquark component(s) of the nucleon and, thus, becomes
suppressed in comparison, say, with $\Delta\to N\pi\,$, though
their phase-spaces are similar.

Such conclusion, because of many uncertainties in the present
calculations, may be still disputable. For example, other authors
differently see the role of 5-quark components in $\Delta$-decays,
to both $N\pi$~\cite{risp} and $N\gamma$~\cite{risg} channels. In
any case, we emphasize that necessity to account for higher
(multiquark) components of conventional baryons opens new
possibilities for understanding the small value of
$\Gamma_{\Theta^+} \,$. Note that accounting for possibility of
the smaller effective $\alpha_s$ in 5-quark configurations, as
discussed above, has not been used in calculations of
Refs.\cite{dpF,cl,dd,cl1}. It may give additional contribution to
the smallness of the $\Gamma_{\Theta^+}\,$.

Now, by analogy, we can assume that configurations with large
number of quarks and antiquarks have a special role also in
production of exotic hadrons. Specifically, we assume that exotics
production goes mainly through many-quark configurations, while
canonical hadrons may be (and, possibly, are) produced mainly
through few-quark configurations.  This should not mean that the
many-quark states hadronize only into exotic multiquark hadrons,
while the few-quark states produce only canonical hadrons. But we
expect that the relative yield of the multiquark hadrons (in
respect to the canonical ones) is higher for many-quark initial
states.

Our hypothesis is a generalization of the suggestion of Karliner
and Lipkin~\cite{kl,lip}, who assumed the $\Theta^+$-production
mainly going through the multiquark resonance. In difference, we
do not stick to particular resonance(s) and admit contributions
from any many-quark state.

\section{Exotics production as a new kind of hard processes}

The known hard processes of hadronic interactions are related, at
the parton level, to the underlying short-time partonic processes.
We can separate all such processes into two groups.

The first, smaller group contains $e^+e^-$ annihilation into
hadrons, heavy quarkonium decays into light hadrons, and also
hadron production in the high-energy $\gamma\gamma$ collisions.
All these processes have no initial light-quarks hadrons. At the
parton level, they start from production of small number (one,
mainly) of (mainly light) quark-antiquark pair(s). Thus,
hadronization in this set of hard processes begins from few-quark
states.

The other, larger group contains such processes as deep inelastic
scattering (DIS), Drell-Yan pair production, high-$p_T$ hadron
production, heavy quark (or, say, $W/Z$) production in hadron
collisions or in photoproduction, and so on. All these processes,
at the particle level, have initially at least one hadron. In the
course of the underlying short-time process, one parton (or small
number of them) becomes knocked out of the hadron(s). The basic
current interest in such hard processes is aimed at the
probability of those partons to be knocked out (structure
functions) and/or at their further evolution (parton interactions
and hadronization).

Meanwhile, these processes, together with the knocked out partons,
produce also remnant(s) of the initial hadron(s), with very many
partons. We have practically no information on structure and later
evolution of such many-parton systems. Indeed, experimentally,
they are mainly lost in the existing detectors. Theoretically,
their description in the framework of QCD has never been achieved
(see, however Ref.\cite{gdkt} for some attempts). This information
vacuum, in respect to hadron remnants, will have to change. Such
systems are very interesting themselves. They may be helpful also
for our present goals, since before hadronization they provide
just a rich source of many-quark states.

Let us discuss the origin of the remnants. In the space-time
picture, higher Fock components of a hadron are known to be
related with short-term fluctuations. Such fluctuations become
averaged out and, thus, are not seen explicitly in long-time-scale
processes (though affect them through radiative corrections). They
are better seen, however, in short-time-scale processes (it is
like to a kind of stroboscopic effects). Thus, hard
(short-time-scale) hadronic processes provide a way to tag
short-term fluctuations with the corresponding characteristic
time. Due to the uncertainty principle, the characteristic number
of partons in a fluctuation is also correlated with the
fluctuation life-time. The shorter is the time-scale of an
underlying process, the higher may be the number of partons in the
fluctuations essential for the process. Only one or several of the
partons are active participant(s) of the ``elementary process''  ,
while all others are spectators, frozen in the hadron remnant. The
number of partons in the remnant is not fixed, but concentrated
near a characteristic number, that increases with decreasing the
time scale of the hard process.

Thus, remnants, that emerge in hard processes, provide many-parton
states. According to our present hypothesis, their hadronization
should generate multiquark hadrons (exotic, in particular) at
higher relative rate, than it is familiar in well-studied
processes. We expect, therefore, that hadron remnants in hard
processes with initial conventional hadrons may appear to be
exotics factories.

This does not mean, however, that they are the only source of
exotics. Fluctuations (with various parton multiplicities) exist
in hadrons independently of one or another process; they
contribute to any process, at least through radiative corrections.
Of course, statistical weight of a particular fluctuation may be
different in different processes. Our hypothesis suggests that
higher-multiplicity fluctuation(s) should stronger contribute to
the production of multiquark hadrons than fluctuations with lower
multiplicity. Thus, a process of exotics production is mainly
related with higher Fock components of initial hadron(s) and,
therefore, tags them. In this sense, exotics production should be
considered as a new kind of hard processes. Such processes contain
short-term fluctuations as a necessary intermediate stage, even if a
particular process, at the hadron level, does not formally contain
any explicit short-time scale.

Close relation between exotics production and many-parton fluctuations
may have observable consequences. For example, if parton multiplicity
of essential configurations is larger for exotics production than for
production of conventional hadrons, then it is natural to expect that
the hadron multiplicity accompanying a produced exotic hadron should
be higher than the average hadron multiplicity.

The above examples demonstrate that our hypothesis is not purely
formal. Even in the present qualitative form, it can be applied to
analyze existing or future experiments.

\section{Experimental $\Theta^+$-production}

Up to now, positive evidences have been published for three exotic
baryons: $\Theta^+(1530),~ \Theta_c^0(3100)$ and
$\Xi_{3/2}^{--}(1860)$ (or $\Phi^{--}(1860)\,$)~\cite{PDG06}. Each
of the two latter states was seen by one group only; they have not
been found in other dedicated experiments. So we will not discuss
them here.

More crucial in the present situation looks the (non)existence of
the $\Theta^+$. The corresponding information (positive or
negative) is much more copious than for any other exotic hadron.
We will not overview all existing publications. However, some of
data, both published and preliminary, seem to touch not only a
problem of the $\Theta^+$-existence, but also a possible mechanism
of the $\Theta^+$-production. It is natural, therefore, to
consider them from the viewpoint of our hypothesis. We will
discuss the positive and negative experimental data separately.

\subsection{Positive data}

a) \textit{ZEUS data.} Let us begin with data of the ZEUS
Collaboration, since they are most advanced in relation to a
mechanism of the $\Theta^+$-production. The Collaboration has
found the $\Theta^+(1530)$ and its antiparticle~\cite{zeus1} in
the kinematical region, where hadroproduction was believed to come
from fragmentation of the knocked-out parton(s)~\cite{exp,kar}.
The proton remnant has been usually considered to escape nearly
unnoticed by the ZEUS detector. In any case, the remnant
contribution in the region used was assumed to be negligible. Such
expectations have been confirmed by measurements of fragmentation
functions for various identified hadrons. In particular,
fragmentation fractions for different charmed hadrons, measured by
ZEUS in this kinematical region~\cite{cop}, coincide with those
measured in $e^+e^-$ annihilation, as was predicted for DIS in the
current fragmentation region (fragmentation of the
knocked-out (anti)quark is expected to be nearly the same as
fragmentation of the (anti)quark produced in $e^+e^-$
annihilation)~\cite{gdkt}.

Special attention was given to production properties of the
$\Lambda(1520)$ and $\Lambda_c^+(2285)$, since they could be
kinematically similar to production of the $\Theta^+(1530)$. It
appears~\cite{zeus2} that the $\Lambda(1520)$ and
$\Lambda_c^+(2285)$ are produced as expected: $\Lambda(1520)$
through fragmentation of the knocked-out quark, while
$\Lambda_c^+(2285)$ through fragmentation of the
$c\overline{c}$-pair, generated by the $\gamma^*$-gluon fusion.

Nevertheless, $\Theta^+(1530)$ in the same kinematical region
clearly demonstrates distributions characteristic for being
generated by hadronization of the target (proton) remnant
(in particular, it is mainly produced in the forward hemisphere,
\textit{i.e.}, at positive pseudorapidity $\eta$)~\cite{zeus2}.
The proton remnant in DIS is a state with superposition of various
many-quark configurations. Thus, the ZEUS data evidently support
our hypothesis of the $\Theta^+$-production coming mainly from
hadronization of such configurations.

Furthermore, our hypothesis leads to expectation that the
$\Theta^+$-production in DIS should change with $Q^2$, because of
changing role of different many-parton configurations.
Experimentally, the $\Theta^+$-yield in ZEUS data decreases with
increasing $Q^2_{\rm min}$ at $Q^2>20$~GeV$^2$~\cite{kar}.
However, it is mainly due to the factor $1/Q^4$ coming from the
photon propagator squared. This factor has no relation to remnant
configurations and can be eliminated, if one considers the
relative yield of the $\Theta^+$ in respect to conventional
hadrons. The ZEUS Collaboration investigated the ratio
$\Theta^+(1530)/\Lambda(1116)$ in the same kinematical region.
This ratio shows slow increase, but, with current large
uncertainties, may be considered also as a constant, about
4\%~\cite{kar}. The situation reminds the case of nucleon
structure functions, which initially looked to be
$Q^2$-independent (Bjorken scaling), but then both theoretical
considerations and more precise measurements have revealed scaling
violation in the structure functions, with slow (logarithmical)
dependence on $Q^2$. Since the underlying physics, the enhanced
role of many-parton configurations at higher $Q^2$, is the same in
these two cases, we expect that the ratio
$\Theta^+(1530)/\Lambda(1116)$ should also logarithmically change
(increase) at high $Q^2$. Evidently, it is a prediction for future
experiments, which can, thus, check our hypothesis.

Meanwhile, we would like to discuss one more feature of the ZEUS
data. The $\Theta^+$ peak is not seen in the $p K_S $ spectrum at
$Q^2>1$~GeV$^2$~\cite{zeus1}. In the HERA kinematics, one has
\mbox{$W^2<10^5$~GeV$^2$,} where $W$ is the $(\gamma^*p)$ c.m.
energy, and the above restriction for $Q^2$ may be rewritten for
the Bjorken variable $x=Q^2/(W^2+Q^2)$ as $x>10^{-5}$. A very
interesting point is that the ZEUS data demonstrate the $\Theta^+$
peak even at $Q^2>1$~GeV$^2$, if one applies additional
restriction \mbox{$W<125$~GeV~\cite{zeus1}, \textit{i.e.},}
\begin{equation} \label{x_B}
x > 6.4\cdot10^{-5}\,.
\end{equation}
It is well known, on the other side, that nucleon structure
functions increase at very small $x$. Therefore, the larger
contributions to the DIS inclusive hadroproduction (in particular,
to the $K_S p$ spectrum) at $Q^2>1$~GeV$^2$ should come from the
lower region
\begin{equation} \label{x_Bl}
10^{-5} < x < 6.4\cdot10^{-5}\,,
\end{equation}
which violates condition (\ref{x_B}). Thus, the ZEUS
data~\cite{zeus1} imply that the $\Theta^+$-production in the
region (\ref{x_Bl}) should not have such enhancement (if any),
which appears at small $x$ for the usual DIS production of the
background non-resonant $K_S p$ system. Note that, kinematically,
the region (\ref{x_Bl}) cannot be reached at HERA in other
$Q^2$-regions investigated by ZEUS (more exactly, at
$Q^2>6$~GeV$^2$). This might explain why observability of the
$\Theta^+$-peak in the ZEUS data does not need any additional
restrictions for $W$ (or $x$) at higher $Q^2$.

The two quantities, $Q^2$ and $x$, are, generally, independent
variables, and a definite value of $x$ may (and, most probably,
does) select some states among various many-parton configurations
with the ``life time'' $\sim 1/\sqrt{Q^2}$, corresponding to the
used value of $Q^2$. If the $\Theta^+$ is not produced indeed at
too small $x$, it could mean, that typical states of the remnant
with presence of a quark having such $x$, may have properties
(\textit{e.g.}, quark energy distributions) that suppress the
$\Theta^+$-formation. Future experiments could study both $Q^2$
and $x$ dependencies of the $\Theta^+$-production in more detail,
to extract interesting information on structure of both the proton
remnant and the $\Theta^+$ itself. The related problems seem to be
worth of more detailed theoretical discussion elsewhere.

b) \textit{Some other positive data.} Manifestations of the
$\Theta^+$-production mechanism are possibly seen also in
preliminary data of the SVD Collaboration. Its higher-statistics
analysis~\cite{svd, svd-m} has confirmed the earlier evidence for
production of the $\Theta^+(1530)\,$ in nucleon-nucleon
collisions~\cite{svd-e}.  Moreover, preliminary data of this
collaboration provide distributions of the $\Theta^+$ in the Feynman
variable $x_F$ and in $p_T$.

The $x_F$-distribution for the $\Theta^+(1530)$ appears to be
very soft: no $\Theta^+(1530)$ is seen at
$|x_F|>0.3$~\cite{svd-m,svd1}, while the observed
$\Lambda(1520)$-yield is much harder, it has maximum at
$|x_F|=0.4\,$~\cite{svd1}. Distribution of the $\Theta^+$ over $p_T$
is also soft (no $\Theta^+$ at $p_T> 1.5$~GeV)~\cite{svd1}.  The two
distributions together imply that the total c.m.-momentum of the
produced $\Theta^+$ is not higher than 1.5~GeV, essentially smaller
than kinematically available (and than that for $\Lambda(1520)$). Such
properties look quite natural in framework of our hypothesis, since we
expect that accompanying hadron multiplicity for the pentaquark
$\Theta^+(1530)$ is higher than, say, for the conventional hyperon
$\Lambda(1520)$. This makes the c.m.-energy available for the
$\Theta^+(1530)$  to be smaller than for the $\Lambda(1520)\,$.

Similar evidence may come also from preliminary data of the HERMES
Collaboration~\cite{lor}. When the standard kinematic constraints,
used for the $\Theta^+$-observation, are appended by the
requirement of an additional detected pion, the signal/background
ratio essentially improves. It reaches 2:1, instead of 1:3 in the
published HERMES data~\cite{herm}. Our hypothesis allows to
understand such an unexpected result. The condition of an
additional pion in the detector enhances the role of events with
higher multiplicity, where, as we argue, the pentaquarks should be
present at higher rate in respect to conventional hadrons than in
average events.

\subsection{Null data}

a) \textit{Low-energy measurements.} Two collaborations have
published impressive null results on the $\Theta^+(1530)\,$ at low
(or relatively low) initial energies.

One of them, the Belle Collaboration, used low-energy kaons,
produced in $e^+e^-$-annihi\-la\-tion, for secondary scattering
inside the detector. Physically, this experiment is similar to
that of \mbox{DIANA}, so they may be directly compared. In both
cases, one studies the charge exchange reaction $K^+n\to pK_S$
inside a nucleus and looks for the $\Theta^+$  as an intermediate
resonance. Experimental results on this charge exchange process
may be directly interpreted in terms of the $\Theta^+$-width. The
Belle data~\cite{bel} do not show any $\Theta^+$-signal and
provide the  restriction
\begin{equation} \label{bel-w}
\Gamma_{\Theta^+} < 0.64~\rm{MeV}\,, \end{equation} which may cast
doubt on the value
\begin{equation} \label{w-ct}
\Gamma_{\Theta^+}= (0.9\pm0.3)~\rm{MeV}\,, \end{equation}
extracted by Cahn and Trilling~\cite{ct} from earlier DIANA
analysis~\cite{di1}. However, that publication was incomplete,
and some additional assumptions on the background contributions
were necessary to extract the value (\ref{w-ct}). More detailed
DIANA data~\cite{bar} allow to find the width with much less
assumptions. Its new value (\ref{expG}) looks extremely
low \footnote{Of course, it is much smaller than the experimental
resolution. The method of extraction of such small width is
described in Ref.\cite{ct}.}, it is not yet quite understood
theoretically, but does not contradict to any other experimental
result on $\Gamma_{\Theta^+}$. In particular, with this new value,
the current Belle data cannot exclude existence of the $\Theta^+$.

The other set of null data is given by the CLAS measurements of
photoproduction off proton and neutron (inside the deutron).
Despite impressive experimental statistics, treatment of several
final states has not revealed any reliable
$\Theta^+$-signal~\cite{clas1, clas2,clas3,clas4}. These results
could pretend to disprove positive evidences of LEPS and SAPHIR.
However, the kinematical region of the CLAS detector cannot
completely overlap those of the two others (the problem is mainly
related with the forward direction), so a more detailed comparison
is necessary for definite conclusions.

Some current theoretical estimations~\cite{kwee,guz} predict even
lower cross sections for the $\Theta^+$-photoproduction, than the
experimental upper limits of CLAS. It is not clear, whether such
expectations may agree with existing positive observations.
However, today's predictions have essential theoretical
uncertainties. They touch, first of all, form factors in exchange
diagrams. The published calculations apply form factors with
properties familiar for conventional hadrons. However, the form
factors might be different if transition, say, $N\to\Theta^+$ goes
mainly in 5-quark configuration. According to ideas of
Feynman~\cite{feyn}, the more constituents has a system, the
faster should change (decrease) its form factor \footnote{The
underlying physics is rather transparent: to change the motion of
a whole system of many constituents one should realize many
interactions between all those constituents, so to drag each of
them.}. Then, according to our hypothesis, photoproduction of the
pentaquark $\Theta^+$ on the nucleon should go as on a 5-quark
system, with a form factor having steeper $t$-dependence than in
photoproduction of conventional hadrons. This might produce the
sharp angular distribution which would make mutually consistent
observation of the $\Theta^+$, \textit{e.g.}, in the forward-looking
detector LEPS and its non-observation in the less-forward detector
CLAS.

If the CLAS data really contain a small $\Theta^+$-signal, consistent
with the published bounds, this signal may be enhanced to the
observable level due to interference effects~\cite{adp}. Thus, the
published CLAS data cannot yet reject existence of the $\Theta^+$.

This is even more true for the COSY-TOF data. The new
measurements~\cite{tof} do not confirm earlier evidence of the same
collaboration for the $\Theta^+$-baryon in the reaction $pp\to pK^0
\Sigma^+$~\cite{tof0}. Instead, they set a strict limit for the
$\Theta^+$-production. However, their new upper boundary for the
cross section is still higher than the theoretical
estimation~\cite{pols} obtained before both new and old analyses.

b) \textit{High-energy measurements.} Null results have been published
also for several types of processes considered to be high-energy ones.
We begin with $e^+e^-\to\,$hadrons (and/or $\gamma\gamma\to\,$hadrons).
Such processes at high energies go mainly through the prompt production
of one quark-antiquark pair, which then hadronizes. This favorably leads
to purely mesonic final states. The events with baryon-antibaryon pairs
need 3 prompt quark-antiquark pairs and provide only small part (say, less
than 1/10) of all high-energy events. Production of pentaquark baryon(s)
needs even more, 5 prompt quark-antiquark pairs, and one may expect that
number of events with pentaquark(s) in high-energy $e^+e^-$ annihilation
is less than 1/10 of events with conventional baryons. Current experimental
data have not reached sufficient precision to observe them at such level.

Similar (and even stronger) conclusions are also true for charmonium
decays, where modes with baryon-antibaryon pairs have the rate of order
1/100 in respect to purely meson modes. This implies that pentaquarks
in charmonium decays may be expected to have rate not more than 1/100
in regard to conventional baryons. Detailed analysis~\cite{as} shows
that the corresponding experimental data have strongly insufficient
precision.

In hadron-hadron collisions, there is the negative search for
the $\Theta^+$ by the SPHINX Collaboration~\cite{sph}. It is usually
claimed to reject the SVD positive result~\cite{svd,svd-m,svd1} (the two
collaborations collected their data at the same Serpukhov proton beam of
70 GeV). However, the SPHINX detector is sensitive mainly to the area
of diffraction dissociation (excitation) of the initial proton. At the
Serpukhov energy, this process can be well described by an effective
pomeron interacting with constituent quarks, without changing the number
of constituents. Therefore, our hypothesis implies strong suppression of
multiquark baryons in the SPHINX experiment. The suppression may be
weaker in the SVD case, where contribution of many-quark fluctuations
could provide a more noticeable effect, than in a special case of
diffraction dissociation.

A large set of high-energy data for the $\Theta^+$-search in
hadron-hadron collisions has been obtained at hadronic colliders.
Their detectors usually tag events with high values of the
transverse energy. Such events are mainly initiated by partons
knocked out from initial hadrons, which then hadronize essentially
as few-parton systems. Multiquark hadrons are very rare in their
hadronization products. For the same events, our hypothesis
suggests to expect a higher rate of exotic hadron production in
remnants of the initial hadrons. However, these remnants are
practically lost for the FNAL detectors CDF and/or D0. The same is
true for the main RHIC detectors, at least, in symmetric collisions
$p+p\,,~Cu+Cu\,$ and $\,Au+Au\,$. Up to now, dedicated
investigation of the hadron remnants has not been planned for those
collider detectors in any approved experiments.

Thus, the current high-energy data, which do not see the
$\Theta^+$, cannot be decisive for the pentaquark problem, even
despite their greater statistics. In framework of our hypothesis,
the reason is that they investigate kinematical regions where
multiquark hadrons are produced with very low rate.

\subsection{Suggestions for further experiments}

Our hypothesis about exotics production mechanism allows us to suggest
new approaches for studying multiquark hadrons. The corresponding
experiments should be oriented to hadrons produced from many-quark
systems. As we explained, a clear example of such systems is
provided by hadron remnants in hard processes with initial hadrons.
At current symmetric configurations of the colliding beams, the
remnants escape into tubes of the storage rings. Thus, to investigate
them, one would need to essentially modify the existing detectors,
to register forward-going hadrons having very high rapidity.

Directions of necessary modification may be demonstrated by the
BRAHMS forward spectrometer. It has recently presented the first
measurements of high-rapidity production of pions, kaons, protons, and
their antiparticles in $pp$-collisions at RHIC ($y_{\,had}\approx3$,
while $y_{\,beam}=5.4$ at $\sqrt{s}=200$~GeV)~\cite{brahms}. The data
for pions and kaons agree with calculations up to Next-To-Leading-Order
in the perturbative QCD, but (anti)protons in that region show yet
unexplained deviations from expected distributions (see
Ref.~\cite{brahms} for more detailed comparison of data \textit{vs.}
theory). Baryon spectroscopy in this region may also appear
non-conventional.

Another approach could be applied, if one used, say, at the FNAL
collider, an asymmetric configuration with $p$ and $\overline p$ beams
of unequal energies.  Then, the Lorentz boost would increase the
angular dimension for the remnant of the less-energetic hadron, so it
could be seen even in the existing detectors, without their
modifications. Such a way would allow to study the kinematical region
in high-energy hadronic collisions, which has not been seen up to now
and which could provide, as we expect, higher rate of exotics
production, than in the published collider experiments.

Note, that the HERA facility is just an asymmetric collider. However,
its asymmetry (the proton momentum much higher than the electron one)
is such that the detectors ZEUS and H1 favor investigations of the
current fragmentation region, while the target fragmentation region,
directly related to the proton remnant, is strongly shrinked. We expect
that the HERA run with less energetic proton beam could allow the
ZEUS detector (and may be H1 as well) to study in more detail the
proton remnant contributions. A very interesting and perspective
direction of studies could be investigation of $x$- and
$Q^2$-dependencies of the $\Theta^+$-production in DIS. In particular,
we expect that, with $Q^2$ growing, the ratio of exotic/conventional
hadron yields should logarithmically increase, in analogy with
logarithnic increase of the small-x hadronic structure functions.

Many-quark systems may emerge also in soft processes, due to short-term
fluctuations of initial hadron(s). However, their relative contribution
is usually rather small and, thus, should provide rather small cross
section for the exotics production as compared to conventional hadron
production \footnote{Note, that hard processes, directly related to the
short-term fluctuations, also have cross sections small in comparison
with soft processes. Separation of a hard process from much more
copious soft background requires some special selection conditions,
\textit{e.g.}, high-$p_T$ or high-$Q^2$ events.}. To increase an
exotics signal/background ratio one needs to apply some selections
enhancing the role of many-parton fluctuations.  A possible way could
be to study events with high (total or charged) hadron multiplicity. Of
course, such events have larger combinatorial background. But we expect
that the relative yield of the exotic $\Theta^+$-baryon \textit{vs.} a
conventional baryon, say, $\Lambda(1520)\,$, should be enhanced for
such events in comparison to the total set of events.  Properties of
the $\Theta^+$-production, as observed by the SVD
Collaboration~\cite{svd,svd-m,svd-e,svd1}, show that also useful for
the exotics observation might be restriction to relatively low energy
of the $(KN)$-pair (as we have explained, this is also related to the
enlarged multiplicity).

The two latter points may be essential in estimating experiments with
null results for exotics searches. Some of those experiments, because
of ``technical'' reasons, apply restrictions on particle energies and/or
on multiplicity of selected events. The above discussion demonstrates
that such restriction can suppress (or enhance!) the exotics
production. Such a possibility should be taken into account in future
experiments.

\section{Summary and conclusions}

Let us summarize the above considerations. We assume that exotic
(and, more generally, multiquark) hadrons are mainly generated
from many-quark partonic configurations, which may emerge either
as short-term fluctuations of initial hadrons in any hadronic
process, or as hadron remnants in hard processes. This implies that
the characteristic time scale for production of multiquark hadrons
is, in any case, shorter than for conventional hadron production.
In this sense, the multiquark production presents a new kind of hard
processes.

We have demonstrated that our assumption is able to give qualitative
explanation why the $\Theta^+$ has been seen in some experiments and
not in others. Moreover, it allows to overcome seeming inconsistencies
between experiments, which are similar at first sight, and to
understand such peculiar features of positive experiments as,
\textit{e.g.}, essential change of signal/background ratio with
relatively small change of registration conditions.

We can also suggest new experiments that might be decisive to check
(and hopefully confirm) existence of the $\Theta^+$ and other multiquark
hadrons. Most important for this purpose could be studies of hadron
remnants in hard processes. Up to now, structure and evolution of those
remnants have not been experimentally investigated and/or theoretically
understood.

If our assumptions appear correct, experiments with production and
investigation of multiquark hadrons could provide new, very interesting
and important information about structure of conventional hadrons and
about properties of their short-term fluctuations. We may expect also,
that studies of multiquark hadrons will reveal both a new hadronic
spectroscopy and new understanding of constituent quarks, with their
properties, masses and couplings, depending on the number of quarks
in the hadron.

\section*{Acknowledgments}

We thank A. Airapetian, S, Chekanov, D. Diakonov, A. Dolgolenko,
M. Karliner, A. Kubarovsky, H. Lipkin, V. Petrov, M. V. Polyakov, and
R. Workman for useful discussions on various sides of the exotics
problem. Ya.A. thanks the Ruhr-Universit\"at-Bochum for hospitality
at some stages of this work.  The work was partly supported by the
Russian State Grant RSGSS-1124.2003.2, by the Russian-German
Collaboration Treaty (RFBR, DFG), by the COSY-Project J\"ulich, by
Verbundforschung ``Hadronen und Kerne'' of the BMBF, by Transregio/SFB
Bonn, Bochum, Giessen of the DFG, by the U.~S.~Department of Energy
Grant DE--FG02--99ER41110, by the Jefferson Laboratory, and by the
Southeastern Universities Research Association under DOE Contract
DE--AC05--84ER40150.


\begin{thebibliography}{99}

\bibitem{GN}
H.~Goldberg and Yu.~Ne'eman, Nuovo\ Cim.\ \textbf{27}, 1 (1963).

\bibitem{GM}
M.~Gell-Mann, Phys.\ Lett.\ \textbf{8}, 214 (1964).

\bibitem{Zw}
G.~Zweig, CERN prepints TH-401, TH-412 (1964).

\bibitem{cool}
R.~L.~Cool \textit{et al.}, Phys.\ Rev.\ Lett.\ \textbf{16}, 1228
(1966); \textbf{17}, 102 (1966).

\bibitem{abr}
R.~J.~Abrams \textit{et al.}, Phys.\ Rev.\ Lett.\ \textbf{19}, 259
(1967).

\bibitem{tys}
J.~Tyson \textit{et al.}, Phys.\ Rev.\ Lett.\ \textbf{19}, 255 (1967).

\bibitem{PDG86}
M.~Aguilar-Benitez \textit{et al.} (Particle Data Group), Phys.\
Lett.\ B\ \textbf{170}, 289 (1986).

\bibitem{jaf1}
R.~L.~Jaffe and K.~Johnson, Phys.\ Lett.\ B\ \textbf{60}, 201 (1976).

\bibitem{jaf2}
R.~L.~Jaffe, invited talk at the \textit{Topical Conf. on Baryon
Resonances, Oxford, July 5-9, 1976}; preprint SLAC-PUB-1774, July
1976; \textit{Oxford\ Top.\ Conf.\ 1976}, p.~455 (QCD161:C45:1976).

\bibitem{jaf3}
R.~L.~Jaffe, Phys.\ Rev.\ D\ \textbf{15}, 267, 281 (1977).

\bibitem{az70}
Ya.~I.~Azimov, Phys.\ Lett.\ B\ \textbf{32}, 499 (1970).

\bibitem{cc}
C.~E.~Carlson, C.~D.~Carone, H.~J.~Kwee, and V.~Nazaryan,
 Phys.\ Rev.\ D\ \textbf{70}, 037501 (2004); hep-ph/0312325.

\bibitem{hys}
J.~S.~Hyslop \textit{et al.}, Phys.\ Rev.\ D\ \textbf{46}, 961 (1992).

\bibitem{dpp}
D.~Diakonov, V.~Petrov, and M.~V.~Polyakov, Z.\ Phys.\ A\ \textbf{359},
305 (1997); hep-ph/9703373.

\bibitem{wk}
H.~Walliser and V.~B.~Kopeliovich, ZhETF\ \textbf{124}, 483 (2003)
[JETP\ \textbf{97}, 433 (2003)]; hep-ph/0304058.

\bibitem{ekp}
J.~Ellis, M.~Karliner, and M.~Praszalovicz, JHEP\ 0405, 002 (2004);
hep-ph/0401127.

\bibitem{PDG04}
S.~Eidelman \textit{et al.} (Particle Data Group), Phys.\ Lett.\ B\
\textbf{592}, 1 (2004).

\bibitem{bur}
V.~Burkert, Int.\ J.\ Mod.\ Phys.\ A\ \textbf{21}, 1764 (2006);
hep-ph/0510309.

\bibitem{PDG06}
W.-M.~Yao \textit{et al.} (Particle Data Group), J.\ Phys.\ G\
\textbf{33}, 1 (2006).

\bibitem{clas1}
M.~Battaglieri \textit{et al.} (CLAS Collaboration), Phys.\ Rev.\
Lett.\ \textbf{96}, 042001 (2006); hep-ex/0510061.

\bibitem{clas2}
B.~McKinnon \textit{et al.} (CLAS Collaboration), Phys.\ Rev.\ Lett.\
\textbf{96}, 212001 (2006); hep-ex/0603028.

\bibitem{clas3}
S.~Niccolai \textit{et al.} (CLAS Collaboration),
Phys.\ Rev.\ Lett.\ \textbf{97}, 032001 (2006); hep-ex/0604047.

\bibitem{clas4}
R.~de~Vita \textit{et al.} (CLAS Collaboration), Phys.\ Rev.\ D\
\textbf{74}, 032001 (2006); hep-ex/0606062.

\bibitem{leps}
T.~Hotta (for the LEPS Collaboration), Acta\ Phys.\ Polon.\ B\
\textbf{36}, 2173 (2005);\\
T.~Nakano (for the LEPS Collaboration), talk at \textit{The Workshop
on $\eta$-photoproduction}, Bochum, Germany, February, 2006;\\
T.~Nakano (for the LEPS Collaboration), talk at the Yukawa
Intern. Seminar (YKIS) 2006, \textit{New Frontiers in
QCD, Exotic Hadrons and Exotic Matter}, Nov.~20 -- Dec.~8, 2006,
Kyoto, Japan~~
$\langle$http://www2.yukawa.kyoto-u.ac.jp/$\sim$ykis06/index.html$\rangle$.

\bibitem{svd}
A.~Aleev, E.~Ardashev, V.~Balandin, S.~Basiladze, \textit{et al.} (SVD
Collaboration), hep-ex/0509033.

\bibitem{svd-m}
A.~Kubarovsky, V.~Popov, and V.~Volkov (for the SVD Collaboration),
in \textit{Proc. of the XXXIII Intern. Conf. on High Energy Physics
(ICHEP'06), Moscow, Russia, July 26 - August 2, 2006}; hep-ex/0610050.

\bibitem{bar}
V.~V.~Barmin, A.~E.~Asratyan, V.~S.~Borisov, C.~Curceanu, \textit{et
al.} (DIANA Collaboration), Yad.\ Fiz.\ \textbf{70}, 39 (2007)
[Phys. At. Nucl. \textbf{70}, 35 (2007)]; hep-ex/0603017.

\bibitem{tr}
Yu.~A.~Troyan, A.~V.~Beljaev, A.~Yu.~Troyan, E.~B.~Plekhanov,
\textit{et al.}, in \textit{Proc. of the 18th Intern. Baldin
Seminar on High Energy Physics Problems: Relativistic Nuclear Physics
and Quantum Chromodynamics (ISHEPP 2006), Dubna, Russia, Sep. 2006};
hep-ex/0610085.

\bibitem{foc}
J.~M.~Link \textit{et al.} (FOCUS Collaboration), Phys.\ Lett.\ B\
\textbf{639}, 604 (2006); hep-ex/0606014.

\bibitem{l3}
P.~Achard, O.~Adrian, M.~Aguilar-Benitez, J.~Alcaraz, \textit{et al.}
(L3 Collaboration), Eur.\ Phys.\ J.\ C\ \textbf{49}, 395 (2007);
hep-ex/0609054.

\bibitem{ank}
M.~Nekipelov, M.~Buescher, M.~Hartmann, I.~Keshelashvili, \textit{et
al.}, J.\ Phys.\ G\ \textbf{34}, 627 (2007); nucl-ex/0610026.

\bibitem{tof}
M.~Abdel-Bary \textit{et al.} (COSY-TOF Collaboration),
hep-ex/0612048.

\bibitem{nom}
O.~Samoylov, D.~Naumov, V.~Cavasinni, P.~Astier, \textit{et al.}
(Nomad Collaboration), Eur.\ Phys.\ J.\ C\ \textbf{49}, 499 (2007);
hep-ex/0612063.

\bibitem{adp}
M.~Amarian, D.~Diakonov, and M.~V.~Polyakov, hep-ph/0612150.

\bibitem{kl}
M.~Karliner and H.~J.~Lipkin, Phys.\ Lett.\ B\ \textbf{597}, 309 (2004);
hep-ph/0405002.

\bibitem{lip}
H.~J.~Lipkin, in \textit{Proceedings of the 32nd International
Conference on High Energy Physics (ICHEP04), Beijing, China,
Aug.~2004}, vol.~\textbf{2}, p.~1033; hep-ph/0501209.

\bibitem{as}
Ya.~Azimov and I.~Strakovsky, Phys.\ Rev.\ C\ \textbf{70}, 035210
(2004); nucl-th/0406312.

\bibitem{hic}
K.~Hicks, in \textit{Proc. of the IX Intern. Conf. on Hypernuclear
and Strange Particle Physics, Mainz, Germany, Oct.~2006};
hep-ph/0703004.

\bibitem{chr1}
C.~V.~Christov \textit{et al.}, in \textit{Proc. of the First
German-Polish Symposium on Particles and Fields, Rydzyna Castle, Poland,
Apr.\ 28 -- May\ 1, 1992}, ed. by H.~D.~Doebner, M.~Pawlowski, and
R.~Raczka, (World Scientific, 1994), p.~3; hep-ph/9303211.

\bibitem{chr2}
C.~V.~Christov \textit{et al.}, Progr.\ Part.\ Nucl.\ Phys.\
\textbf{37}, 91 (1996); hep-ph/9604441.

\bibitem{dia}
D.~Diakonov, in \textit{Lectures at the Advanced Summer School on
Non-Perturbative Quantum Field Theory, Peniscola, Spain, June 2-6,
1997}, p.~1; hep-ph/9802298.

\bibitem{dp}
D.~Diakonov and V.~Petrov, in \textit{The Frontier of Particle
Physics/Handbook of QCD; Boris Ioffe Festschrift}, ed. by
M.~Shifman, (World Scientific, 2001) vol.~1, p.~359;
hep-ph/0009006.

\bibitem{pra}
M.~Praszalowicz, Acta\ Phys.\ Polon.\ B\ \textbf{35}, 1625 (2004);
hep-ph/0402038.

\bibitem{arn}
R.~A.~Arndt \textit{et al.}, Phys.\ Rev.\ C\ \textbf{68}, 042201 (2003);
Erratum, C\ \textbf{69}, 019901 (2004); nucl-th/0308012.

\bibitem{ct}
R.~N.~Cahn and G.~H.~Trilling, Phys.\ Rev.\ D\ \textbf{69}, 011501
(2004); hep-ph/0311245.

\bibitem{gib}
W.~R.~Gibbs, Phys.\ Rev.\ C\ \textbf{70}, 045208 (2004);
nucl-th/0405024.

\bibitem{sib}
A.~Sibirtsev \textit{et al.}, Phys.\ Lett.\ B\ \textbf{599}, 230 (2004);
hep-ph/0405099.

\bibitem{mia}
G.-S.~Yang, H.-Ch.~Kim, and K.~Goeke, Phys.\ Rev.\ D\ \textbf{75},
094004 (2007) ; hep-ph/0701168.

\bibitem{lip1}
H.~Lipkin, hep-ph/0703190.

\bibitem{z&s}
Ya.~B.~Zeldovich and A.~D.~Sakharov, Yad.\ Fiz.\ \textbf{4}, 394 (1966)
[Sov.\ J.\ Nucl.\ Phys.\ \textbf{4}, 283 (1967)];\\
Ya.~B.~Zeldovich and A.~D.~Sakharov, Acta\ Phys.\ Hungar.\ \textbf{22},
153 (1967).

\bibitem{dpF}
D.~Diakonov and V.~Petrov, Phys.\ Rev.\ D\ \textbf{72}, 074009 (2005);
hep-ph/0505201.

\bibitem{cl}
C.~Lorce, Phys.\ Rev.\ D\ \textbf{74}, 054019 (2006); hep-ph/0603231.

\bibitem{dd}
D.~Diakonov, Write-up of the talks at \textit{Quarks-2006
(St.~Petersburg, Russia, May, 2006)} and at \textit{Quark Confinement
and Hadron Spectrum VII (Ponta Delgada, Sep.~2006)}; AIP\ Conf.\ Proc.\
\textbf{892}, 258 (2007); hep-ph/0610166.

\bibitem{cl1}
C.~Lorce, arXiv:0705.1505 [hep-ph].

\bibitem{risp}
Q.~B.~Li and D.~O.~Riska, Phys.\ Rev.\ C\ \textbf{73}, 035201 (2006);
nucl-th/0507008.

\bibitem{risg}
Q.~B.~Li and D.~O.~Riska, Nucl.\ Phys.\ A\ \textbf{766}, 172 (2006);
nucl-th/0511053.

\bibitem{gdkt}
Yu.~Dokshitzer, L.~Gribov, V.~Khoze, and S.~Troyan, Phys.\ Lett.\ B\
\textbf{202}, 276 (1988).\\
L.~V.~Gribov, Yu.~L.~Dokshitzer, S.~I.~Troyan, and V.~A.~Khoze,
ZhETF\ \textbf{94}, 12 (1988) [Sov.\ Phys.\ JETP\ \textbf{67}, 1303
(1988)].\\ A.~V.~Anisovich, G.~Anzivino, F.~Arzarello, Ya.~I.~Azimov,
\textit{et al.}, Nuovo\ Cim.\  A\ \textbf{106}, 547 (1993).

\bibitem{zeus1}
S. Chekanov \textit{et al.} (ZEUS Collaboration), Phys.\ Lett.\ B\
\textbf{591}, 7 (2004); hep-ex/0403051.

\bibitem{exp}
S.~Chekanov (for the ZEUS Collaboration), in \textit{Proc. of the XII
Intern. Workshop on Deep Inelastic Scattering (DIS 2004), Strbske
Pleso, High Tatras, Slovakia, April, 2004}, ed. D.~Bruncko,
J.~Ferencei, and P.~Strizenec, p.~579; hep-ex/0405013.

\bibitem{kar}
U.~Karshon (for the ZEUS Collaboration), in \textit{Proc. of the
Intern. Workshop PENTAQUARK04, SPring-8, Hyogo, Japan, 20-23 July
2004}, p.~75; hep-ex/0410029.

\bibitem{cop}
N.~Coppola (on behalf of the H1 and ZEUS Collaborations), in
\textit{Proc. of the 41st Rencontres de Moriond - QCD and high energy
hadronic interactions, La Thuile, France, March, 2006}; hep-ex/0605056.

\bibitem{zeus2}
S.~Chekanov (for the H1 and ZEUS Collaborations), in \textit{Proc. of
the Intern. Europhysics Conf. on High-Energy Physics (July 21-27,
2005) Lisboa, Portugal}; PoS (HEP 2005) 086 (2006); hep-ex/0510057.

\bibitem{svd-e}
A.~Aleev, N.~Amaglobeli, E.~Ardashev, V.~Balandin, \textit{et al.}
(SVD Collaboration), Yad.\ Fiz.\ \textbf{68}, 1012 (2005) [Phys.\ At.\
Nucl.\ \textbf{68}, 974 (2005)]; hep-ex/0401024.

\bibitem{svd1}
P.~Ermolov (for the SVD Collaboration), talk at the seminar of IHEP,
April 2006.

\bibitem{lor}
W.~Lorenzon (for the HERMES Collaboration), in \textit{Proc. of the
Intern. Workshop PENTAQUARK04, SPring-8, Hyogo, Japan, 20-23 July,
2004}, p.~66; hep-ex/0411027.

\bibitem{herm}
A.~Airapetian \textit{et al.} (HERMES Collaboration), Phys.\ Lett.\
B\ \textbf{585}, 213 (2004); hep-ex/0312044.

\bibitem{bel}
K.~Abe \textit{et al.} (Belle Collaboration), Phys.\ Lett.\ B\
\textbf{632}, 173 (2006);  hep-ex/0507014.

\bibitem{di1}
V.~V.~Barmin, V.~S.~Borisov, G.~V.~Davidenko, A.~G.~Dolgolenko,
\textit{et al.} (DIANA Collaboration), Yad.\ Fiz.\ \textbf{66}, 1763
(2003) [Phys.\ At.\ Nucl.\ \textbf{66}, 1715 (2003)]; hep-ex/0304040.

\bibitem{kwee}
H.~Kwee, M.~Guidal, M.~V.~Polyakov, and M.~Vanderhaeghen,
Phys.\ Rev.\ D\ \textbf{72}, 054012 (2005); hep-ph/0507180.

\bibitem{guz}
V.~Guzey,  hep-ph/0608129.

\bibitem{feyn}
R.~P.~Feynman, \textit{Photon-Hadron Interactions}, (W.~A.~Benjamin,
Inc., 1972).

\bibitem{tof0}
M.~Abdel-Bary \textit{et al.} (COSY-TOF Collaboration), Phys.\
Lett.\ B\ \textbf{595}, 127 (2004); hep-ex/0403011.

\bibitem{pols}
M.~V.~Polyakov, A.~Sibirtsev, K.~Tsushima, W.~Cassing, and K.~Goeke,
Eur.\ Phys.\ J.\ A\ \textbf{9}, 115 (2000); nucl-th/9909048.

\bibitem{sph}
Yu.~M.~Antipov, A.~V.~Artamonov,~V.~A.~Batarin, O.~V.~Eroshin,
\textit{et al.} (SPHINX Collaboration), Eur.\ Phys.\ J.\ A\ \textbf{21},
455 (2004); hep-ex/0407026.

\bibitem{brahms}
I.~Arsene, I.~G.~Beardon, D.~Beavis, S.~Bekele, \textit{et al.}
(BRAHMS Collaboration), accepted for publication in Phys.\ Rev.\ Lett.;
hep-ex/0701041.

\end{thebibliography}
\end{document}